# $La_{14.33}(CN_3)_8N$, a Nitridocarbonate based on the $(CN_3)^{5-}$ Ion

Markus Ströbele,[a] Mykhailo Brytskyi,[b] Carl P. Romao,[c] Scott Kroeker,[b] and Hans-Jürgen Meyer*[a]

ABSTRACT: Metal nitridocarbonates containing the $(NCN)^{2-}$ ion are established for a multitude of metal atoms throughout the periodic table. The preparation of nitridocarbonate compounds containing a $(CN_3)^{5-}$, $(C_2N_4)^{6-}$, or $(C_3N_6)^{6-}$ ion remains challenging, although corresponding species have been established. The new compound $La_{14.33}(CN_3)_8N$ was synthesized in a conventional solid-state metathesis reaction from a mixture of $LaF_3$, $Na_2(CN_2)$ and $Li_3N$ in an open system at 550 °C. The reaction conditions, including the formation and de-composition of $La_{14.33}(CN_3)_8N$ are analyzed on basis of thermoanalytic studies. The crystal structure of $La_{14.33}(CN_3)_8N$ was refined by single-crystal X-ray diffraction, and the mid-IR frequencies resulting from two crystallographically distinct $(CN_3)^{5-}$ ions in the structure are reported. $^{23}Na$, $^{7}Li$ and $^{13}C$ NMR studies provide additional support for the identity of the compound. The crystal structure of $La_{14.33}(CN_3)_8N$ is isotypic to that of $Ce_{15}(BN_3)_8N$.

## INTRODUCTION

Over the past decades, an increasing number of dinitridocarbonates have been synthesized which are commonly assigned as carbodiimides or cyanamides. These compounds encompass a wide variety of metal cations, including rare-earth carbodiimides of the type $RE_2(CN_2)_3$ with $RE$ = Sc – Lu (except Pm), transition metal carbodiimides such as $Zr(CN_2)_2$, $Hf(CN_2)_2$, $Cr_2(CN_2)_3$, $M(CN_2)$ with $M$ = Mn, Fe, Co, Ni,[1-5] as well as p-block carbodiimides like $In_{2.24}(CN_2)_3$, $Bi_2(CN_2)_3$, $M(CN_2)$ with $M$ = Pb, Zn, Cd, Sn.[6-11] Additionally, tetracyanamidometallates containing the $[M(CN_2)_4]^{n-}$ anion (with $M$ = Si, Ge, and other $M$ atoms) have been reported.[12-17] In all these structures, the $(NCN)^{2-}$ ion is present, although other anionic species based on the elements carbon and nitrogen could exist beside $(NCN)^{2-}$ and $(CN)^{-}$.[18-20] For instance, various compositions of nitridoborate anions have been characterized with five distinct compositions, as $(BN)^{n-}$, $(NBN)^{3-}$, $(BN_3)^{6-}$, $(B_2N_4)^{8-}$, and $(B_3N_6)^{9-}$ units, all shown in Figure 1.[21] These anions represent the cyanide analogue $(BN)^{n-}$, the dinitridocarbonate analogue $(NBN)^{3-}$, the carbonate analogue trinitridoborate $(BN_3)^{5-}$, the $(B_3N_6)^{9-}$ ion whose structure corresponds to a fully dehydrogenated melamine $(C_3N_6)^{6-}$, and the oxalate analogue $(B_2N_4)^{8-}$, all shown in Figure 1.

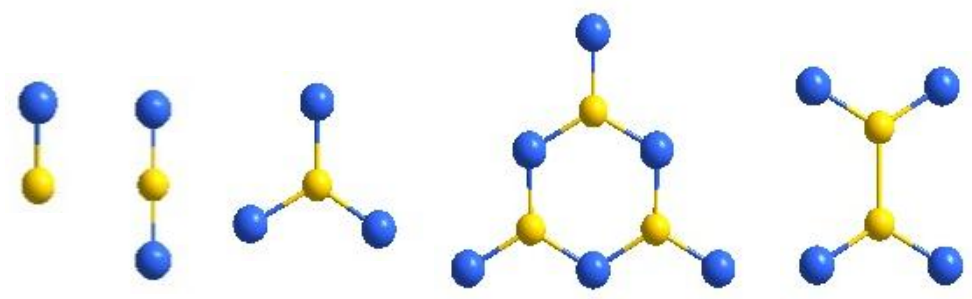

**Figure 1.** Structures of reported nitridoborate anions $(BN)^{n-}$, $(BN_2)^{3-}$, $(BN_3)^{6-}$, $(B_3N_6)^{9-}$, and $(B_2N_4)^{8-}$ with boron atoms shown yellow and N atoms blue.

Both classes of compounds, nitridoborates and nitridocarbonates, can be readily synthesized via solid-state metathesis (SSM) reactions[22] involving a metal halide and either lithium carbodiimide or lithium nitridoborate as precusrsors. For $RE$ = La, the reactions exhibit ignition temperatures near 460 °C for reaction (1) and near 555 °C for reaction (2):

$$2\ RECl_3 + 3\ Li_2(CN_2) \rightarrow RE_2(CN_2)_3 + 6\ LiCl \qquad (1)^{23}$$

$$RECl_3 + Li_3(BN_2) \rightarrow RE(BN_2) + 3\ LiCl \qquad (2)^{22}$$

In spite of the apparent similarity of products given in reactions (1) and (2), these SSM reactions produce compounds with distinct anion structures. The crystal structure of $RE_2(CN_2)_3$ ($RE$ = Sc-Lu, except Pm)[24-26] contains a carbodiimide ion, $(N{=}C{=}N)^{2-}$, while the structure of $RE(BN_2)$ is better expressed as $RE_3(B_3N_6)$ (RE = La, Ce, Pr, Nd, Sm, Gd)[21, 27, 28] because it contains the cyclic $(B_3N_3)^{9-}$ ion (see Figure 1), formed through cyclotrimerization of $(BN_2)^{3-}$ ions. Notably, a corresponding cyclotrimerization of $(CN_2)^{2-}$ ions has not yet been reported in SSM reactions with $Li_2(CN_2)$. However, the fully dehydrogenated melaminate

[a] Section for Solid State and Theoretical Inorganic Chemistry, Institute of Inorganic Chemistry, Auf der Morgenstelle 18, Eberhard Karls Universität Tübingen, D-72076 Tübingen, Germany.
[b] Department of Chemistry, University of Manitoba, Winnipeg, R3T 2N2, Canada.
[c] Department of Materials, Faculty of Nuclear Sciences and Physical Engineering, Trojanova 13, Czech Technical University in Prague, 120 00 Prague, Czech Republic.

$Pb_3(C_3N_6)$ has been recently reported, synthesized under high-pressure conditions from elements.[29]

In addition, partially dehydrogenated melaminates have been synthesized by solid-state metathesis reactions. One such example is the compound $Cu_3(C_3N_6H_3)$,[30, 31] obtained from a reaction between CuCl and $Na(HCN_2)$ via cyclotrimerization of $(HNCN)^-$ ions. Analogous reactions of $MCl_2$ with molten K(OCN) have shown to involve the cyclotrimerization of $(OCN)^-$ ions to yield compounds of the type $M_3(O_3C_3N_3)_2$ ($M$ = Ca, Sr, Eu), which show strong second harmonic generation (SHG) response.[32-35]

These examples demonstrate that cyclotrimerization reactions are commonly obtained from 16-electron anions, exemplified by $(NBN)^{3-}$ and $(OCN)^-$.

Carbonate analogues, such as $(BN_3)^{6-}$ and $(CN_3)^{5-}$, appear to be less common and, when accessible, are typically obtained in low yields. Nitridoborates containing the $(BN_3)^{6-}$ ion are evidenced in $La_2(BN_3)$,[28] $La_3(BN_3)N$, $La_5(BN_3)(B_3N_6)$, and $La_6(BN_3)(B_3N_6)N$, all obtained from SSM reactions using appropriate ratios of $LaCl_3$, $Li_3(BN_2)$, and $Li_3N$.[21] $La_6(BN_3)O_6$ was prepared correspondingly, with LaOCl serving as an oxygen source.[36] Moreover, the metal-rich compound $La_3(B_2N_4)$ involves a dimerization of $(NBN)^{3-}$ ions in a classical SSM reaction between $LaCl_3$ and $Li_3(BN_3)$ under reducing conditions (with Li).[28, 37]

A number of compounds featuring the trinitridocarbonate ion $(CN_3)^{5-}$ have been recently synthesized by means of high-pressure chemistry. The compounds $Sb(CN_3)$,[38] and $Ln_3O_2(CN_3)$ with $Ln$ = La, Eu, Tb, Ho, Yb[39] were synthesized using laser heating in diamond anvil cells under high nitrogen pressure conditions (25 – 54 GPa). These reactions were apparently designed to generate nitride phases by reacting Sb or partly oxidized lanthanides ($Ln$) with nitrogen. The presence of carbon atoms in the discovered compounds is assumed to originate from the diamond anvils used in the experiments. The compounds were described as being recoverable at ambient conditions.

Laser heating experiments at even higher pressures (90–111 GPa) and temperatures up to 2500 K have yielded the compounds $LaCN_3$, $TbCN_3$, $CeCN_5$, and $TbCN_5$.[40] Their high-pressure crystal structures are based on single-bonded $CN_4$ tetrahedra which are interconnected via N–N bridges and shared $CN_4$ tetrahedra to form three-dimensional network structures. These compounds have not been verified to exist under ambient conditions. In addition, the $(C_2N_5)^{7-}$ ion has been found in $Eu_4Fe_{0.864(6)}(C_2N_5)_2$ by laser heating of $Eu(N_3)_2$, $EuC_2$ and Fe in a diamond anvil cell at 50(3) GPa.[41]

These results, concerning the formation of compounds containing $(CN_3)^{5-}$, $(C_3N_6)^{3-}$ and $(C_2N_5)^{7-}$ ions as well as extended framework structures containing $CN_3$ and $CN_4$ units, are not entirely unexpected under high pressure conditions, since both the coordination number and the degree of linkage are generally considered to increase with increasing pressure. However, all of these compounds have so far been obtained only in small quantities. This raises the question of whether such compounds can also be synthesized as bulk materials by conventional solid-state reactions.

Classical solid-state syntheses have indeed yielded crystalline compounds containing the trinitridocarbonate ion, namely $(Sr_9N_{1.33})SrIn_3(CN_3)$ and $Sr_4\ (Sr_6N)_2In_4(CN_3)_4$.[42] These were obtained from reactions of strontium, indium, $Sr_2N$, and $NaN_3$ in nickel ampoules using a sodium flux at 550 °C. The carbon source required for the formation of the $(CN_3)^{5-}$ ions was proposed to originate from the nickel-based reaction vessel. Notably, the synthesis was carried out under elevated $N_2$ pressure, generated by the decomposition of $NaN_3$ in a sealed reaction vessel. However, the resulting pressure is not considered to be the decisive factor for the successful formation of these compounds.

Taken together, these high-pressure syntheses and conventional solid-state reactions undoubtedly represent important advances in the chemistry of nitridocarbonate compounds. Nevertheless, the development of a reliable, high-yield synthesis based on a well-defined carbon nitride precursor and carried out under controlled solid-state reaction conditions remains highly desirable.

The synthesis of compounds containing the trinitridocarbonate ion has previously been investigated by means of solid-state reactions between carbodiimides and nitrides. However, under given conditions, the formation of the trinitridocarbonate ion was not observed. Instead, carbodiimide nitrides were formed, as illustrated in the reaction equations (3) and (4) with $RE$ = La, Ce.

$$2\ RECl_3 + Li_2(CN_2) + Li_3N \rightarrow RE_2Cl(CN_2)N + 5\ LiCl \qquad (3)^{43}$$

$$3\ RECl_3 + 3\ Li_2(CN_2) + Li_3N \rightarrow RE_3(CN_2)_3N + 9\ LiCl \qquad (4)^{44}$$

In this work, we investigate a solid-state metathesis (SSM) reaction that is closely related to reaction (4), which unexpectedly leads to the formation of $La_{14.33}(CN_3)_8N$, a compound containing trinitridocarbonate ions. Its crystal structure is isostructural with that of $Ce_{15}(BN_3)_8N$, thereby revealing a striking structural analogy between nitridocarbonate and nitridoborate chemistries.

## Results and Discussion

### Synthesis and Crystal Structure

$La_{14.33}(CN_3)_8N$ was prepared by heating a mixture of lanthanum fluoride, $Na_2(CN_2)$, and $Li_3N$ from room temperature to 550 °C under an inert atmosphere, with no holding period required. Prolonged heating at 550 °C or variation of the cooling rate affected only the crystallinity of the product. The reaction yielded dark, rod-shaped plate-like crystals of $La_{14.33}(CN_3)_8N$ together with the metathesis byproducts LiF and NaF, as well as minor amounts of $La_2O_2(CN_2)$[45] (see Experimental Section).

The crystal structure of $La_{14.33}(CN_3)_8N$ was determined from single-crystal X-ray diffraction data and refined in the centrosymmetric space group $R\bar{3}c$. The lattice parameters and selected crystallographic data are summarized in Table 1.

Table 1. Lattice parameters and selected data based on single-crystal refinement at 250 K.

| | |
|---|---|
| Formula | $La_{14.33}(CN_3)_8N$ |
| Formula weight | 2468.86 |
| CCDC No. | 2512450 |
| Space group | $R\bar{3}c$ |
| Crystal system | trigonal |
| Wavelength /Å | 0.71073 |
| $a$/Å | 14.4385(3) |
| $b$/Å | 14.4385(3) |
| $c$/Å | 21.5664(6) |
| $\theta$/° | 120 |
| Cell volume /Å$^3$ | 3893.6(2) |
| $Z$ | 6 |
| Density (calculated) /g/cm$^3$ | 6.317 |
| Absorption coefficient /mm$^{-1}$ | 23.424 |
| Theta range for data collection /° | 2.494 to 25.343 |
| Reflections collected | 28560 |
| Independent reflections | 797 [$R_{int}$ = 0.0214] |
| Completeness | 100 % to theta = 25.242° |
| Absorption correction | Gaussian |
| Data / restraints / parameters | 797 / 6 / 76 |
| GooF | 1.589 |
| $R$-values [I >2$\sigma$(I)] | $R_1$ = 0.0148; $wR_2$ = 0.0337 |
| $R$-values (all data) | $R_1$ = 0.0148; $wR_2$ = 0.0337 |
| Largest diff. peak and hole | 0.648 and -0.808 e.Å$^{-3}$ |

The crystal structure contains three crystallographically distinct lanthanum sites. Two of them, La1 and La2, occupy general positions (36f), whereas La3 is located on the special 18e position with site symmetry 2. La1 is nine-coordinate and bonded to nitrogen atoms of the trinitridocarbonate ions as well as to the nitride ion (N5), present in the structure. The La1–N bond lengths range from 2.490(5) to 2.964(5) Å (La1–N3 = 2.490(5) and 2.929(5) Å; La1–N5 = 2.6628(3) Å). In contrast, La2 and La3 are both eight-coordinate, with La2–N distances ranging from 2.394(5) to 2.815(5) Å and La3–N distances from 2.482(5) to 2.789(5) Å. The coordination environments of all three lanthanum sites are shown in Figure 2.

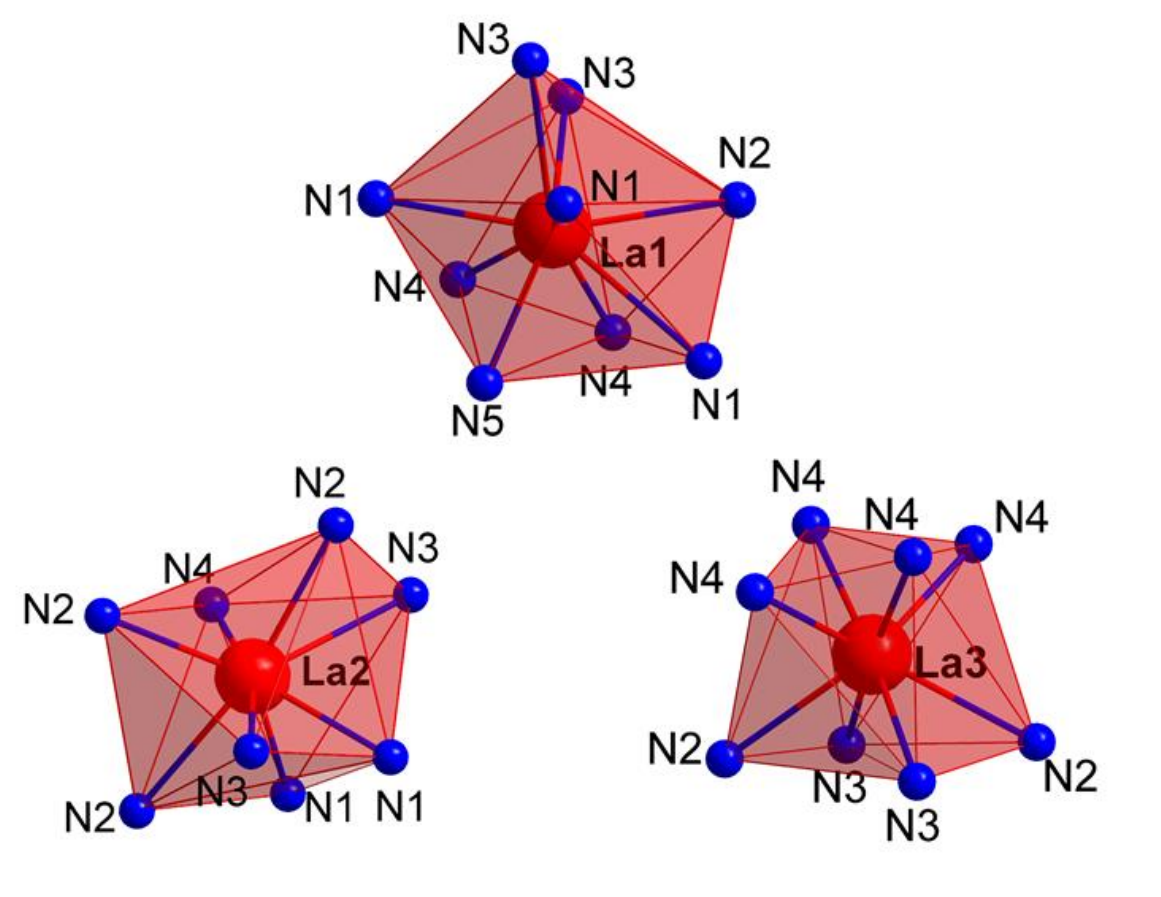

**Figure 2.** Coordination polyhedra of the lanthanum sites in $La_{14.33}(CN_3)_8N$. Lanthanum atoms are shown in red; nitrogen atoms in blue.

The isolated nitride ion (N5) occupies the 6b Wyckoff position on the threefold rotoinversion axis and is octahedrally coordinated by six lanthanum ions. The La1–N5 distance of 2.6628(3) Å agrees well with the nearest-neighbor La–N distance in rock-salt-type LaN. Using the experimental lattice parameter of a = 5.298(2) Å at 148 K, the latter is calculated as d(La–N) = a/2 = 2.649 Å, demonstrating the close similarity of the La–N bonding in both compounds.

The site occupation factor (SOF) of La3 was refined to 0.853(3), thus revealing a deficiency on this site. Furthermore, two distinct $(CN_3)^{5-}$ ions are present in the structure, of which one (based on C1) is situated on a general (36f) position with C–N distances ranging from 1.372(8) to 1.404(7) Å, and the other one is situated on a special (12c) position with its carbon atom (C2) on a threefold rotation axis, revealing three equivalent C–N distances at 1.374(5) Å. These distances are in full agreement with previously reported trinitridocarbonate compounds.[38-40, 42] The surrounding of the $(CN_3)^{5-}$ ion with lanthanum ions is most characteristic, following the motif of a tricapped trigonal-prism, displayed in Figure 3.

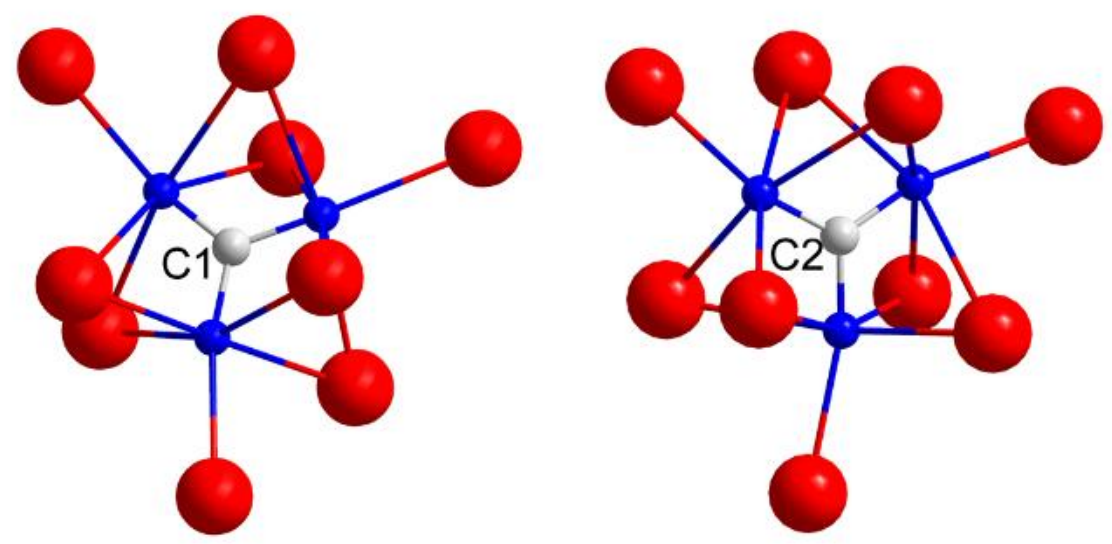

**Figure 3**. Coordination environments of the two distinct $(CN_3)^{5-}$ units, centered with C1 and C2 atoms. Lanthanum atoms are shown in red; nitrogen atoms in blue, and carbon atoms in white.

The overall crystal structure of $La_{14.33}(CN_3)_8N$, shown in Figure 4, can be considered isostructural with that of $Ce_{15}(BN_3)_8N$,[46] which has been described as a mixed valence compound corresponding to $Ce^{(4+)}{}_6Ce^{(3+)}{}_9(BN_3{}^{(6-)})_8N^{(3-)}$. The agreement of the atomic positions in both structures is shown in a comparison in Table S1. Consistent with their isostructural relationship, the two frameworks realize charge neutrality by different cationic schemes. In $Ce_{15}(BN_3)_8N$, the anionic charge of the planar $(BN_3)^{6-}$ units is compensated internally by a mixed-valent cerium sublattice ($Ce^{4+}/Ce^{3+}$). In $La_{14.33}(CN_3)_8N$, all lanthanum is trivalent; because $(CN_3)^{5-}$ carries one unit negative charge less than $(BN_3)^{6-}$, overall neutrality is achieved by a slight deficiency on the lanthanum sublattice. This is reflected crystallographically by the partial occupancy of the La3 site, which results in an overall composition $La_{14.33}(CN_3)_8N$. Thus, while both phases preserve the same polyhedral connectivity, via N-$Ln_6$ octahedra linked by planar trigonal [$XN_3$] groups ($X$ = B or C), they differ in how the cation sublattice balances the anionic framework: a small number of lanthanum vacancies in the lanthanum compound versus mixed lanthanide valence in the cerium analogue.

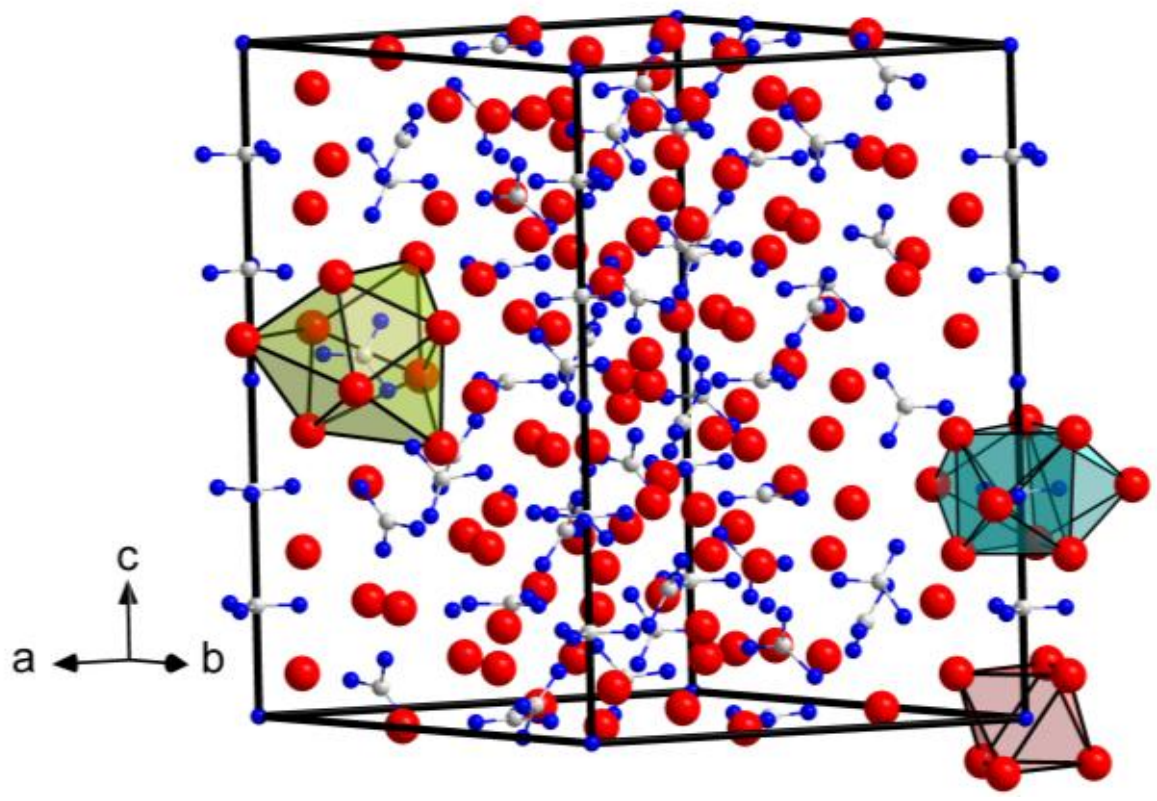

**Figure 4.** Unit cell of $La_{14.33}(CN_3)_8N$, with characteristic coordination polyhedra surrounding the two distinct $(CN_3)^{5-}$ units and the octahedral coordination environment of $N^{3-}$ ion. La is shown red, carbon white, and nitrogen blue.

### Thermal Analysis and Synthetic Considerations

The implementation of a solid-state synthesis method for a given system requires an assignment and optimization of reaction conditions with respect to the composition, temperature, and duration. This usually requires a number of experiments, especially if a pure product is desired. Thermal analysis can be very helpful, especially for those solid-state reactions that are running with a sufficiently high reaction rate. This is true for SSM reactions, which are exothermic reactions that ignite at a certain temperature. Typical ignition temperatures for the preparation of metal nitridocarbonates by SSM reactions (see eq. (1), (3), and (4)) range between 400 and 600 °C.

The synthesis of $La_{14.33}(CN_3)_8N$ was accomplished from a stoichiometric mixture of $LaF_3$, $Na_2(CN_2)$, and $Li_3N$ in a metal container under inert gas conditions (see Experimental Section). A differential scanning calorimetry (DSC) measurement was performed the same way, using gold plated metal containers, provided with a seal. The DSC of the formation reaction of $La_{14.33}(CN_3)_8N$ is shown in Figure 5, revealing an ignition, represented by a strongly exothermic effect at about 470 °C. The determined energy of the ignition is -460 J/g which makes up 11 % of the energy release of TNT (-4184 J/g).[47, 48]

The preparation of metal carbodiimides is most commonly performed with $Li_2(CN_2)$, which can be easily prepared in high yield and purity. However, as we have previously mentioned, the synthesis of $La_{14.33}(CN_3)_8N$ was established with $Na_2(CN_2)$. Therefore, we were prompted to investigate the thermochemical behaviour of $Na_2(CN_2)$ in some more detail, shown in the DTA of $Na_2(CN_2)$ in the inset in Figure 5. The melting of $Na_2(CN_2)$ is obtained with an onset at 620 °C, followed by recrystallization on cooling. The intermediate formation of mixed $LiNa(CN_2)$[49] could lower the melting point even more (the melting point onset of $LiNa(CN_2)$ is at 514 °C).

In general, the formation of a molten phase considerably accelerates solid-state reactions by enhancing mass transport. In exothermic solid-state metathesis (SSM) reactions, the heat released at the ignition point may be sufficient to generate such a transient molten state. This appears to be the case for the present reaction, as the formation of crystalline $La_{14.33}(CN_3)_8N$ occurs essentially instantaneously and does not require any subsequent annealing, in contrast to conventional diffusion-controlled solid-state reactions.

The appearance of a molten state, right at the ignition point of a SSM reaction, has been seen in previous studies of the trimerization of cyanate into cyanurate in the reaction mixture with $MCl_2$, where K(OCN) acts as a melt and reactant below its melting temperature (melting point depression), best demonstrated for $M_3(C_3N_3O_3)_2$ compounds with $M$ = Ca, Ba.[35, 50] In the same way, the extremely high exothermicity related to the formation reaction of $La_{14.33}(CN_3)_8N$ makes an early melting of $Na_2(CN_2)$ inevitable, whereby the (endothermic) melting effect is fully overcompensated by the exothermic reaction.

**Figure 5.** Differential scanning calorimetry (DSC) of the heating process of a mixture based on appropriate proportions of $LaF_3$, $Na_2(CN_2)$, and $Li_3N$ with the exothermic formation of $La_{14.33}(CN_3)_8N$. A corresponding DSC of the melting and recrystallization of $Na_2(CN_2)$ is shown in the inset. Both plots are corrected for background noise.

The thermal stability of $La_{14.33}(CN_3)_8N$ was investigated by simultaneous differential thermal analysis and thermogravimetry (DTA/TG). The measurements reveal the onset of decomposition slightly below 600 °C (Figure 6), yielding LaN as the predominant crystalline decomposition product together with additional phases of lower crystallinity.

Taken together with the synthesis experiments, these results indicate that the temperature window for the preparation of $La_{14.33}(CN_3)_8N$ by solid-state metathesis is limited to approximately 500–570 °C. This finding once again illustrates that the stability range of a compound between its formation and decomposition may be remarkably narrow. Consequently, new compounds can easily be overlooked in empirically guided solid-state syntheses if the reaction conditions are not carefully

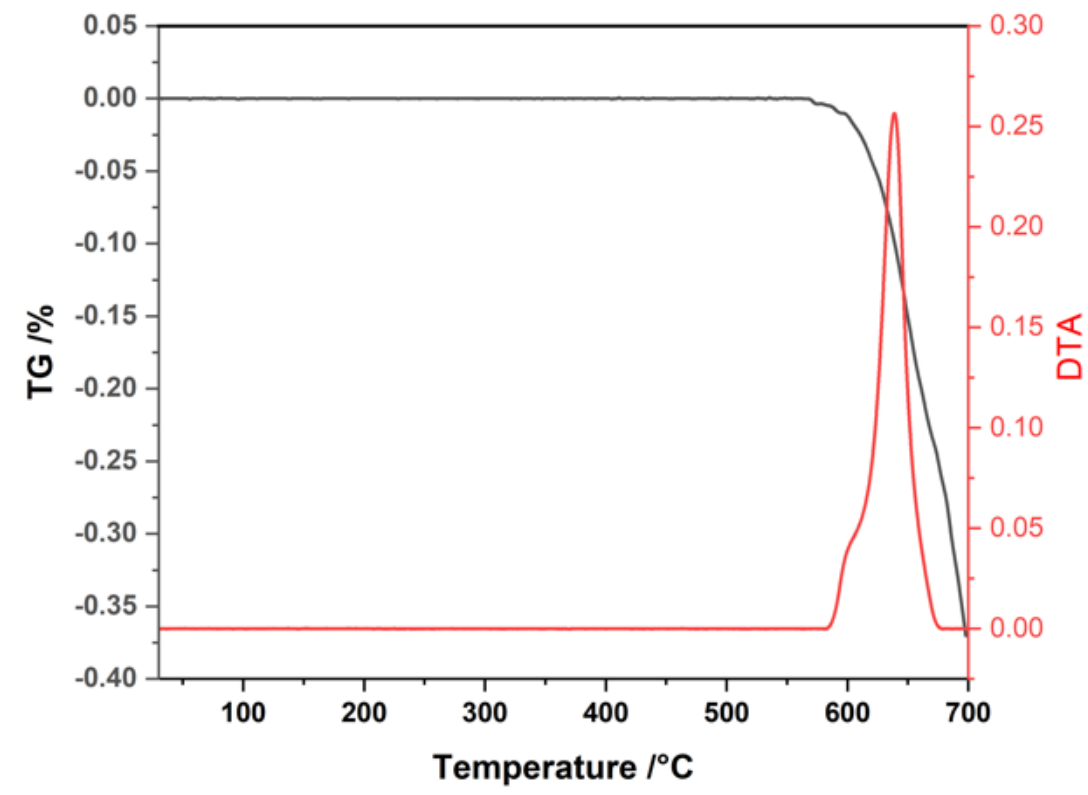

optimized.[51]

**Figure 6.** Decomposition of $La_{14.33}(CN_3)_8N$ monitored by TG/DTA measurement, representing the mass loss (black line) related with the simultaneous endothermic response (red line) with an onset below 600 °C. Both plots are corrected for background noise.

### Infrared (IR) Spectroscopy

Mid-IR spectra of trigonal oxyanions such as carbonate and nitrate are generally characterized by four fundamental vibrational modes: the symmetric stretching vibration ($\nu_1$) at 1040–1105 cm$^{-1}$, the out-of-plane bending vibration ($\nu_2$) at 810–906 cm$^{-1}$, the asymmetric stretching vibration ($\nu_3$) at 1275–1590 cm$^{-1}$, and the in-plane bending vibration ($\nu_4$) at 670–756 cm$^{-1}$. Of these, $\nu_2$, $\nu_3$, and $\nu_4$ are IR-active, whereas the symmetric stretching mode $\nu_1$ is IR-forbidden for an ideal trigonal-planar ion.[52] Within the isotypic series $(NO_3)^-$, $(CO_3)^{2-}$, $(BO_3)^{3-}$, $(CN_3)^{5-}$, and $(BN_3)^{6-}$, increasing bond lengths are accompanied by a continuous shift of the characteristic $\nu_3$ stretching vibration toward lower wavenumbers.

The crystal structure of $La_{14.33}(CN_3)_8N$ contains two crystallographically distinct trinitridocarbonate ions. One anion (C1) occupies a general position, reducing the local symmetry and lifting the degeneracy of the vibrational modes. Consequently, the $\nu_1$ vibration becomes weakly IR-active and the degenerate modes are expected to split. The second trinitridocarbonate ion (C2) is located on a threefold rotation axis, preserving the higher local symmetry.

Consistent with these structural features, the IR spectrum of $La_{14.33}(CN_3)_8N$ exhibits intense absorption bands between 1000 and 1250 cm$^{-1}$ (Figure 7). The strongest bands, observed at approximately 1200 cm$^{-1}$ and 1100 cm$^{-1}$, are assigned predominantly to the asymmetric stretching ($\nu_3$) vibrations of the two crystallographically distinct $(CN_3)^{5-}$ ions, present in a ratio of 1:3. Compared with the corresponding $\nu_3$ vibration of the carbonate ion, these bands are shifted to lower wavenumbers, consistent with the longer, less polar, and therefore weaker C–N bonds relative to the C–O bonds in carbonate. Additional absorption bands arising from the minor side phase $La_2O_2(CN_2)$ are also observed and are identified in Figure S1.[45]

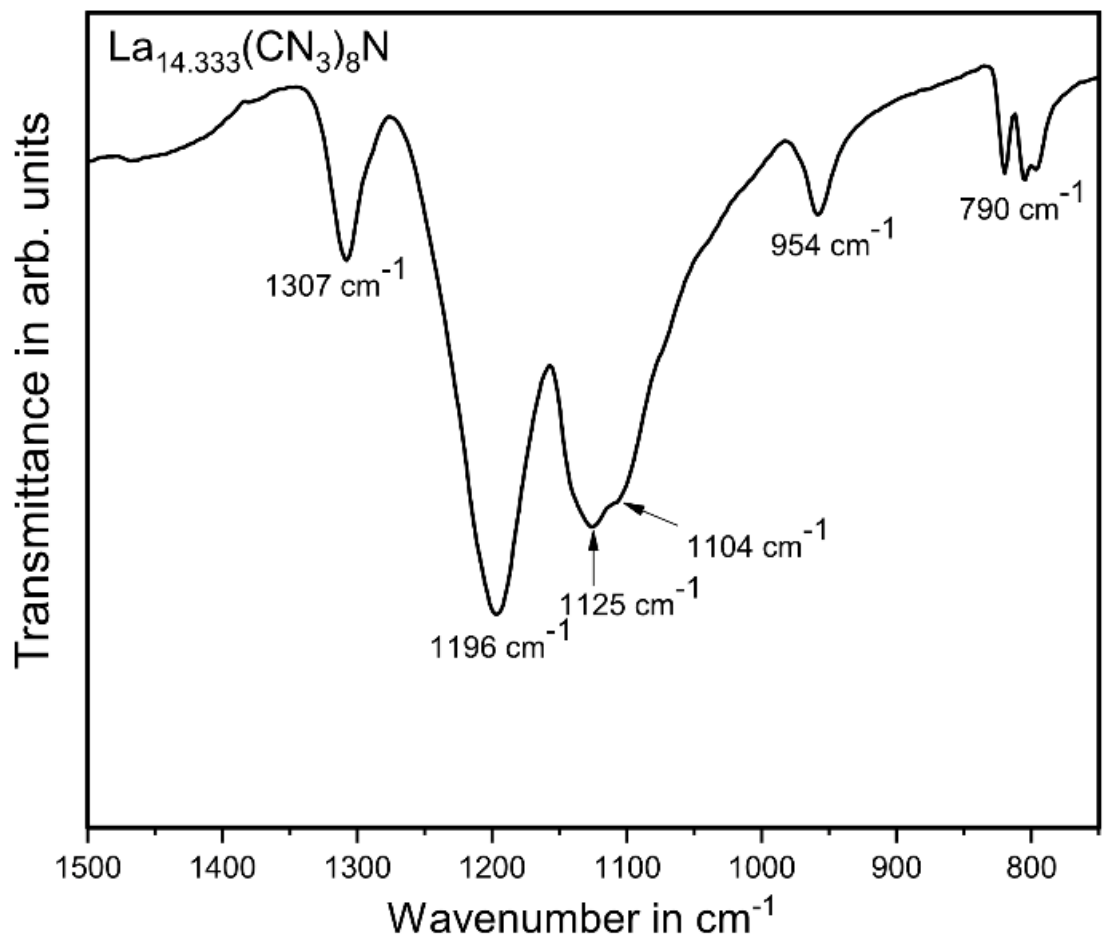

**Figure 7.** Mid IR modes of trinitridocarbonate units recorded for $La_{14.33}(CN_3)_8N$.

### Nuclear magnetic resonance (NMR) spectroscopy

The presence of LiF and NaF in the product is confirmed by $^{19}F$ MAS NMR (Figure 8), with NaF appearing at -201 ppm[53] and LiF at -221 ppm[54]. $^{7}Li$ MAS NMR has one sharp peak at -1.0 ppm[55] (Figure S2) and $^{23}Na$ MAS NMR likewise shows only a single narrow peak at +7 ppm at two applied magnetic fields[56, 57] (Figure S3 for 11.7 T spectrum; 14.1 T spectrum not shown). No other signals were observed in the $^{19}F$, $^{7}Li$ or $^{23}Na$ MAS NMR spectra.

$^{13}C$ MAS NMR spectra of the title compound and a reference compound, $La_2O_2(CN_2)$, are shown in Figure 9. Despite the poor signal-to-noise ratio due to low $^{13}C$ natural abundance, long relaxation times, and interference from probe background, three prominent signals are observed for the $La_{14.33}(CN_3)_8N$ product mixture with approximately equal intensities: +182 ppm, +166 ppm and +132 ppm. The peak at +166 ppm aligns with the largest signal in the $La_2O_2(CN_2)$ spectrum, confirming powder x-ray diffraction evidence that this side phase is present in the sample mixture. However, both spectra show similar signals around +135 and +156 ppm, despite the absence of other carbon-bearing phases in the x-ray diffractogram. These peaks are likely due to poorly crystalline graphite (+130 ppm)[58] or graphitic carbon nitride formed during the reaction (+155 ppm)[59] , neither of which would be expected to appear in x-ray diffraction but which could be responsible for the black appearance of the sample. The remaining peak at +182 ppm is assigned to the $CN_3^{5-}$ species in $La_{14.33}(CN_3)_8N$. This signal is more deshielded than related N-bound carbons (+160-175 ppm)[60, 61], however $^{13}C$ chemical shifts as high as +185 ppm were calculated for melem carbons in corrugated g-$C_3N_4$ models[62]. The two chemically inequivalent $CN_3^{5-}$ in $La_{14.33}(CN_3)_8N$ are structurally similar and likely to be unresolved within the broad (*ca.* 5 ppm) peak centered at +182 ppm. To our knowledge, this is the first report of the $^{13}C$ NMR chemical shift of $CN_3^{5-}$.

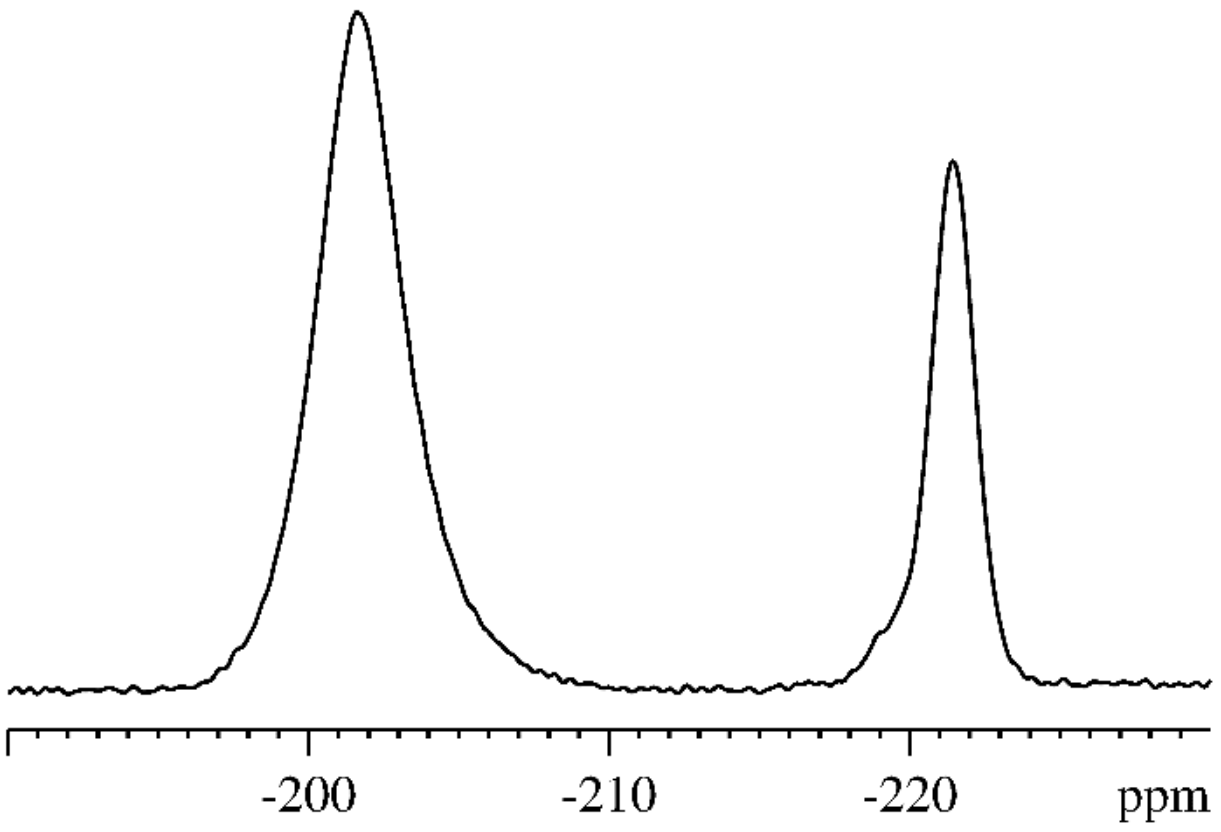

**Figure 8.** $^{19}F$ MAS NMR spectrum of $La_{14.33}(CN_3)_8N$.

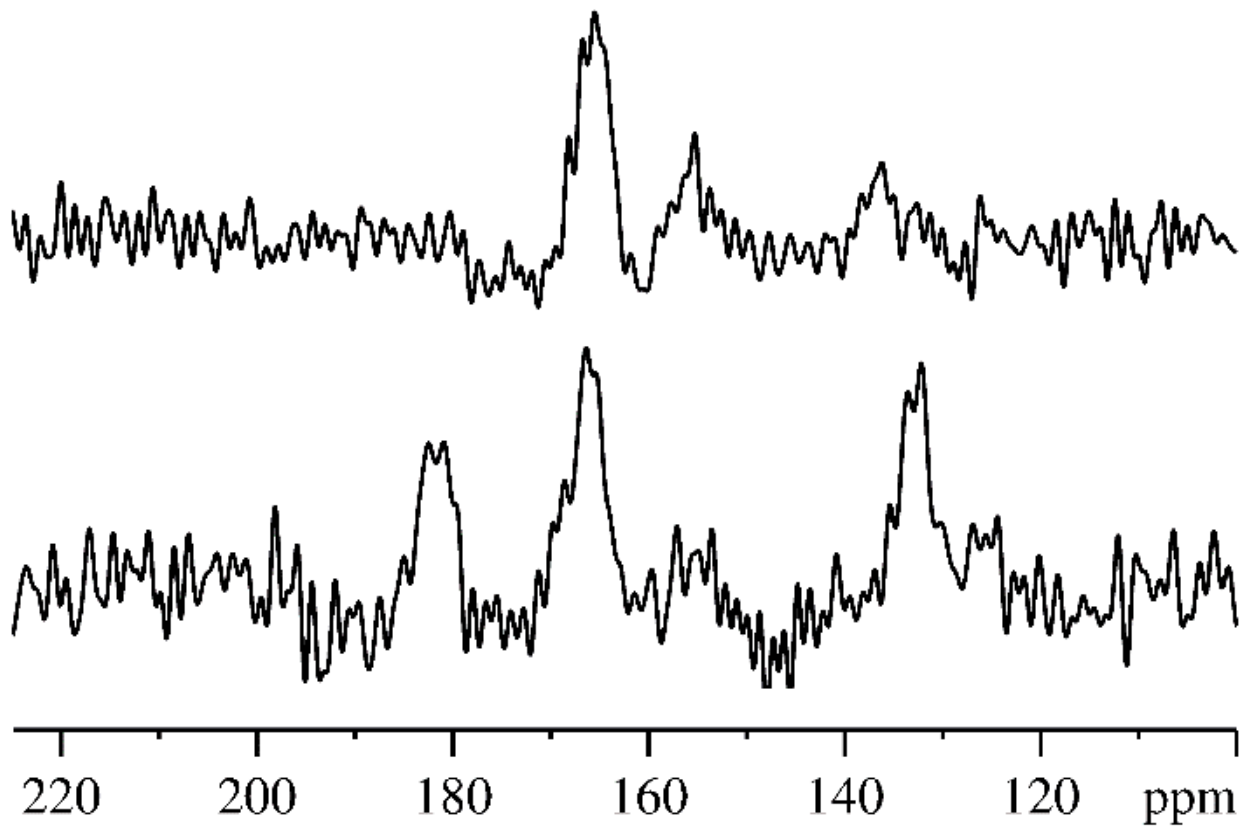

**Figure 9.** $^{13}C$ MAS NMR of $La_2O_2(CN_2)$ (top) and $La_{14.33}(CN_3)_8N$ (bottom). Note that neither sample is pure, but each contains additional phases. See text for details

## Density functional theory (DFT)

The vibrational properties of $La_{14.33}(CN_3)_8N$ were investigated further using DFT calculations of its optic phonon energies and eigenvectors. Due to the large size of the unit cell, it was not practical to include the La vacancy in the calculations, and therefore they were performed using the fully occupied $La_{15}(CN_3)N$ structure. The electronic structure of $La_{15}(CN_3)N$ was found to be closed-shell with a band gap of 0.5 eV (allowing LO-TO splitting). The phonon energies calculated for $La_{15}(CN_3)N$ show good agreement with the experimental results (Figure 7). These findings validate the use of the fully occupied structure to model $La_{14.33}(CN_3)_8N$. The calculated phonon density of states and the mode effective charges (MEC) of the optic modes are shown in Figure 10. The MEC is a measure of the dipole moment induced by the atomic displacements and therefore can be used as a measure of the IR activity of the modes. The large number of atoms in the unit cell leads to a multiplicity of phonons within each observed IR band. However, using the MEC as guidance, we assign the observed mid-IR bands to four types of trinitridocarbonate vibrations, which are shown schematically in Figure 10. The modes above 600 cm$^{-1}$ involve distortions of the $(CN_3)^{5-}$ units, whereas those below 400 cm$^{-1}$ involve motions of the La cations in conjunction with librations of trinitridocarbonates. Within the continuum of modes below 400 cm$^{-1}$, modes with lower energy involve progressively larger displacements of the La cations relative to those of the $(CN_3)^{5-}$ units.

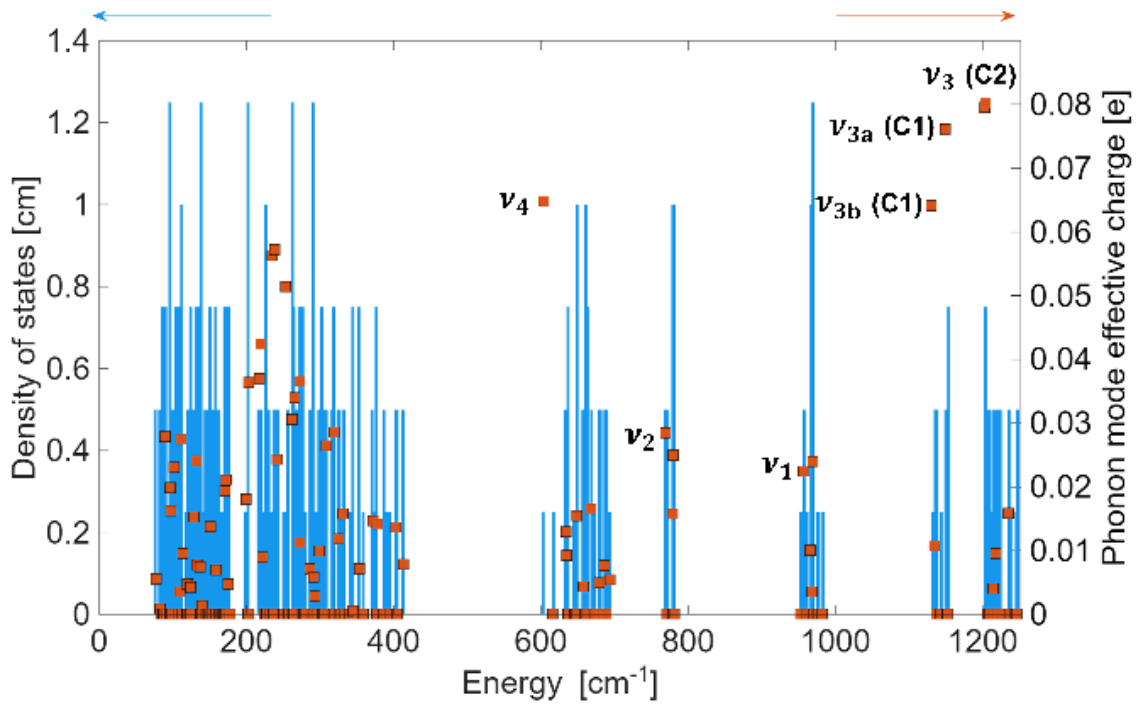

**Figure 10:** Calculated phonon density of states (blue histogram) and mode effective charges (orange squares) of $La_{15}(CN_3)_8N$. Several modes with high IR activity are assigned as symmetric stretches ($\nu_1$), out-of-plane bending ($\nu_2$), asymmetric stretches ($\nu_3$), and in-plane bending ($\nu_4$).

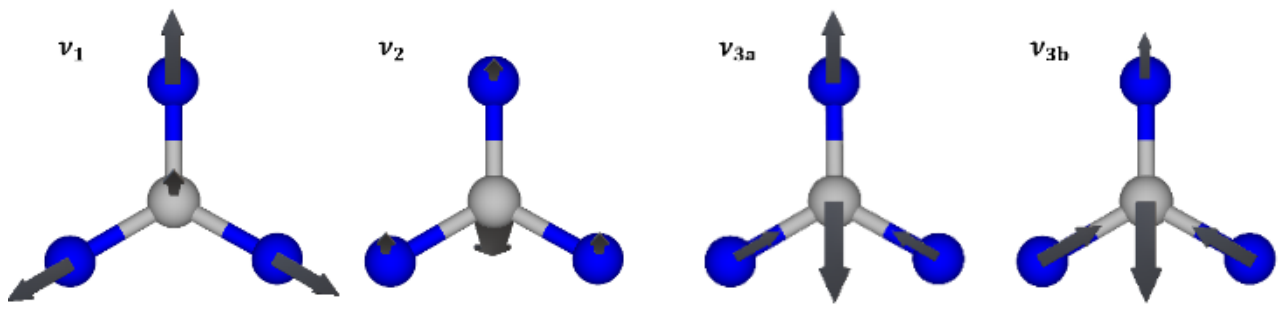

**Figure 11:** Atomic displacements within trinitridocarbonate units corresponding to the modes observed in the mid-IR spectrum of $La_{14.33}(CN_3)_8N$ (see Figure 7). The $\nu_1$ mode appears at 954 cm$^{-1}$ and the $\nu_2$ mode at 790 cm$^{-1}$. The $\nu_3$ mode involving the C2 site is observed at 1196 cm$^{-1}$, while that involving the C1 site creates two peaks, at 1125 and 1104 cm$^{-1}$, based on the relative magnitudes of the displacement of the nitrogen atoms.

The vibrational modes in the mid-IR region show a low degree of eigenvector mixing between the C1 and C2 sites, *i.e.*, the modes can be separated into those where one site moves and the other is stationary. For the symmetric $\nu_1$ and $\nu_2$ modes, the difference in energy between the motions of the C1 and C2 sites is small enough that no splitting can be observed experimentally. However, the asymmetric stretches show a significant difference, with the $\nu_3$ band of the more symmetric C2 site appearing 50 cm$^{-1}$ above that of the C1 site. The $\nu_3$ mode of the C1 site further splits by 20 cm$^{-1}$ depending on the relative displacements of the N atoms, creating two observable peaks labeled $\nu_{3a}$ and $\nu_{3b}$. The appearance of multiple peaks in the asymmetric stretching region therefore can be used as a probe of the crystallographic environment around the $(CN_3)^{5-}$ units (Figure 10).

## Conclusion

Traditionally, nitridoborates and nitridocarbonates have been synthesized via solid-state metathesis (SSM) reactions departing from a metal chloride and $Li_2(CN_2)$, the most-common carbodiimide source. However, variations in these reactants, particularly in the lithium and chloride components, can significantly influence the outcome of a reaction. In this context, $Na_2(CN_2)$ emerges as a promising alternative reagent for SSM reactions aimed at developing new carbodiimide compounds. Nevertheless, its synthesis of $Na_2(CN_2)$ needs to be improved for higher purity, as has already been accomplished for $Li_2(CN_2)$ and $K_2(CN_2)$[24]. The employment of $Na_2(CN_2)$ in the synthesis of $La_{14.33}(CN_3)_8N$ enables a reactive intermediate state, most likely through the formation of a molten state facilitating the product formation, including the formation of the $(CN_3)^{5-}$ ion. A comparable behavior has been observed in the Na flux mediated syntheses of $(CN_3)^{5-}$-containing compounds in the In-Sr-C-N system, as well as in cyclotrimerization reactions leading to $M_3(C_3N_3O_3)$ compounds.

Future studies will show whether the formation or intentional introduction of a melt is a beneficial strategy for synthesizing compounds containing trinitridocarbonate or other new compounds containing $(CN_x)^{n-}$ anions. Ultimately, these findings may open new pathways for a broader class of compounds featuring the $(CN_3)^{5-}$ ion, which is isoelectronic with $(BN_3)^{6-}$, $(BO_3)^{3-}$, $(CO_3)^{2-}$, and $(NO_3)^{-}$.

## Experimental Section

Manipulations of starting materials, such as charging the reaction container with starting materials, as well as products were performed in an argon-filled glove box (MBraun, labmaster 130, $O_2$ < 1 ppm, $H_2O$ < 1 ppm).

### Synthesis of $La_{14.33}(CN_3)_8N$

Stoichiometric amounts of $LaF_3$ (Sigma-Aldrich, anhydrous, 99.99 %), $Na_2(CN_2)$, and $Li_3N$ (Thermo Scientific Chemicals, 99.4 %) were weighed in an argon-filled glovebox with a molar ratio 43:24:27, corresponding to the reaction

$$43\ LaF_3 + 24\ Na_2(CN_2) + 27\ Li_3N \rightarrow La_{43}(CN_3)_{24}N_3 + 48\ NaF + 81\ LiF$$

The resulting reaction mixtures were transferred into niobium ampoules.[63] The reactions were carried out in a muffle furnace in a Schlenk tube with a length of 40 cm and an inner diameter of 35 mm, closed with a GL45 cap equipped with a silicon seal (Schott AG, Mainz, Germany) at 550 °C for 6 h with heating and cooling rates of 2 K $min^{-1}$. After cooling to room temperature, the reaction products were recovered in the glovebox and gently ground. The samples were obtained as black microcrystalline powders containing $La_{14.33}(CN_3)_8N$ (estimated 90 % with respect to La) together with NaF, LiF, $La_2O_2(CN_2)$ as secondary phases (Figure S4). No further washing or purification steps were performed in order to avoid decomposition of the nitride/cyanamide framework.

Due to electroneutrality $La_{14.33}(CN_3)_8N$ should be a white powder. As the product was obtained as black powder, the presence of x-ray amorphous carbon was assumed and confirmed by $^{13}C$-NMR.

### Synthesis of $Na_2(CN_2)$

The synthesis of $Na_2(CN_2)$ was carried out in a Schlenk tube with a length of 40 cm and an inner diameter of 35 mm, closed with a GL45 cap equipped with a silicon seal (Schott AG, Mainz, Germany). The starting materials NaH and $Na(HCN_2)$ were mixed in a 1:1 molar ratio and homogenized in an agate mortar, inside a glove box under argon atmosphere. The mixture was transferred into an open Pt vial and subsequently placed into the Schlenk tube. The synthesis was carried out analogous to a previously reported procedure.[63] The tube was heated in a Simon–Müller furnace at 450 °C with heating and cooling rates of 2 K $min^{-1}$ and a holding time of 30 min. The reaction yielded a white powder containing $Na_2(CN_2)$ (estimated 85% according to PXRD) with $Na_5H(CN_2)_3$ (10 %)[64] and $Na_5(CN)(CN_2)_2$[19] (5%) as side phases.

### Powder X-ray diffraction

All reaction products were investigated by powder X-ray diffraction (PXRD) using a StadiP diffractometer (Stoe, Darmstadt) with Ge-monochromated Cu-$K_{\alpha 1}$ radiation, and a Mythen1 Detector.

### Single-Crystal X-ray Diffraction

Data collections were performed on a Rigaku XtaLAB Synergy-S single-crystal X-ray diffractometer equipped with HyPix-6000HE detector and monochromated Mo-$K_\alpha$ radiation ($\lambda$ = 0.7107 Å) at 250 K. X-ray intensities were corrected for absorption with a numerical method (crystal faces) using CrysAlisPro 1.171.41.67a (Rigaku Oxford Diffraction, 2024)[65]. This software was also used to calculate reconstructed reciprocal space sections. The average structures were solved by direct methods (SHELXT)[66] and refined by full-matrix least-squares methods performed with SHELXL-2015[67] as implemented in Olex2 1.5-ac5-024.[68]

CCDC 2512450 contains the supplementary crystallographic data for $La_{14.33}(CN_3)_8N$. These data can be obtained free of charge from The Cambridge Crystallographic Data Centre via www.ccdc.cam.ac.uk/structures.

### Nuclear magnetic resonance (NMR) spectroscopy

Samples for solid-state NMR were packed into magic-angle-spinning (MAS) rotors in a glove box under argon to ensure an air- and moisture-free environment.

$^7Li$, $^{23}Na$ and $^{19}F$ MAS NMR direct-polarization experiments were conducted on a Bruker Avance III 500 MHz ($B_0$ = 11.74 T) spectrometer using 1.3 mm MAS rotors spinning at 55 kHz. $^7Li$ spectra were collected with a small tip angle of 22.5° (0.1875 µs), recycle delays of 1 and 60 s (to ensure full relaxation), 1024 coadded transients, and referenced to a 1.0 M LiCl solution at 0.00 ppm. $^{23}Na$ spectra were collected with a short tip-angle pulse of 0.25 µs, a recycle delays of 5 s, 64 coadded transients, and referenced to a 1.0 M NaCl solution at 0.00 ppm. $^{19}F$ spectra

used a 1.5 µs pulse, a 240 s recycle delay, 16 coadded transients, and were externally referenced to NaF(s) at -221 ppm.

$^{13}C$ and $^{23}Na$ MAS NMR single-pulse experiments were performed on a Varian INOVA 600 MHz ($B_0$ = 14.1 T) using a 3.2 mm MAS rotor spinning at 12 kHz. $^{13}C$ direct-polarization spectra were collected using a 90° pulse (5 µs) with a recycle delay of 120 s, 2000 coadded transients, and externally referenced to adamantane at +38.4 ppm. Given the inherently low NMR sensitivity of these samples, it was not possible to measure spin-lattice relaxation times, with the consequence that the relative intensities of peaks representing different phases may not be quantitatively accurate. $^{13}C$ MAS spectra were processed using backward linear prediction to remove the probe background signal. $^{23}Na$ spectra were collected with a short tip-angle pulse of 0.31 µs, a recycle delay of 5 s, and 256 coadded transients. Attempts to acquire an $^{15}N$ MAS NMR spectrum were unsuccessful, due to the low $^{15}N$ natural abundance and presumably long relaxation times.

#### Thermoanalytic studies

Differential scanning calorimetry (DSC) was performed using a DSC 204 F1 Phoenix instrument (Netzsch, Selb, Germany). The starting materials were sealed into 100 µL gold-plated (5 µm) steel autoclaves (Bächler Feintech AG, Hölstein, Switzerland) under inert gas conditions. The formation of $La_{14.33}(CN_3)_8N$ was investigated from room temperature to 550 °C, applying a heating rate of 2 °C/min. The DSC was calibrated in accordance with the manufacturer's instructions using high-purity CsCl, $Ag_2SO_4$, $RbNO_3$, Benzoic acid, $KNO_3$, biphenyl, and elementary Ga.

Differential thermal and thermogravimetric analysis (DTA/TG) of the decomposition of $La_{14.33}(CN_3)_8N$ and for the analysis of the melting behavior of $Na_2(CN_2)$ were performed using a Netzsch Jupiter STA 449 F3 thermal analyzer with heating (cooling) rate of 2 °C/min. The samples were measured in an open corundum crucible under continuous argon flow.

#### Infrared Spectroscopy

Infrared spectra were recorded using a VERTEX 70 FT-IR Spectrometer in the range of 400 $cm^{-1}$ to 4000 $cm^{-1}$ using KBr pellets of $La_{14.33}(CN_3)_8N$. Samples were prepared in a glove-box under Argon atmosphere.

#### Density Functional Theory

DFT calculations were performed using the Abinit software package (v. 10).[69] The GGA-PBE exchange–correlation functional[70] was used with the dispersion correction of Grimme.[71] The PAW method[72] was used with a plane-wave basis set energy cutoff of 24 Ha outside of the PAW spheres and 100 Ha inside them. A Monkhorst–Pack grid[73] of **k** points with real-space basis vectors [−2 2 2], [2 −2 2], [2 2 −2] was used to sample reciprocal space for the calculation of the electronic structure. The phonon energies and eigenvectors were calculated using density functional perturbation theory (DFPT) at the Γ (**q** = (0 0 0)) point. Structural relaxation to an internal pressure of 0.5 MPa was performed prior to the DFPT calculation. All computational data are publicly available from Doi:10.5281/zenodo.18012468.

## Conflicts of interest

There are no conflicts to declare.

## Data availability

Crystallographic data have been deposited at the CCDC 2512450 ($La_{14.33}(CN_3)_8N$). Data are available within the article. The data that support the findings of this study are available on request from the corresponding author, H.-J. Meyer.

## Acknowledgements

This research was generously supported be the Deutsche Forschungsgemeinschaft (DFG) through the grant ME 914/34-1. SK is grateful for support from the Natural Sciences and Engineering Research Council (NSERC) of Canada and the Canada Foundation for Innovation (CFI). Computational resources were provided through bwHPC and the German Research Foundation (DFG) through grant no INST 40/467-1 FUGG (JUSTUS cluster). C.P.R. acknowledges support from the project FerrMion of the Ministry of Education, Youth and Sports, Czech Republic, co-funded by the European Union (CZ.02.01.01/00/22_008/0004591).

TOC

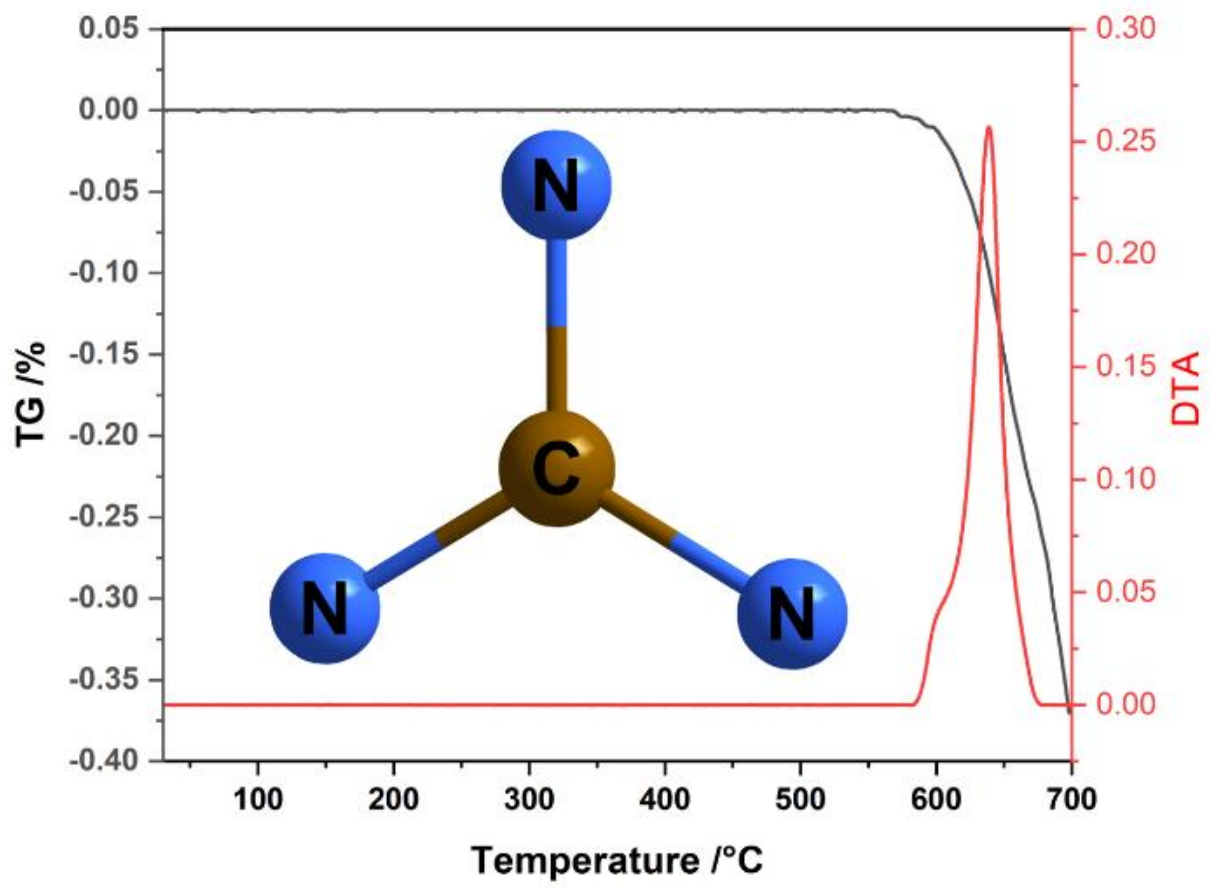

Thermal stability of the trinitridocarbonate $La_{14.33}(CN_3)_8N$, recorded by differential thermal analysis (DTA) and thermogravimetry (TG) in an open system.

Supporting Information for

# $La_{14.33}(CN_3)_8N$, a Nitridocarbonate based on the $(CN_3)^{5-}$ Ion

Markus Ströbele,[a] Mykhailo Brytskyi,[b] Carl P. Romao,[c] Scott Kroeker,[b] and Hans-Jürgen Meyer*[a]

[a] *Section for Solid State and Theoretical Inorganic Chemistry, Institute of Inorganic Chemistry, Auf der Morgenstelle 18, Eberhard Karls Universität Tübingen, D-72076 Tübingen, Germany.*

[b] *Department of Chemistry, University of Manitoba, Winnipeg, R3T 2N2, Canada.*

[c] *Department of Materials, Faculty of Nuclear Sciences and Physical Engineering, Trojanova 13, Czech Technical University in Prague, 120 00 Prague, Czech Republic.*

**Table of contents**

Table S1: Comparison of atomic position in $La_{14.333}(CN_3)_8N$ and $Ce_{15}(BN_3)_8N$.

| | | $La_{14.333}(CN_3)_8N$ | | | | $Ce_{15}(BN_3)_8N$ | | |
|---|---|---|---|---|---|---|---|---|
| Wyckoff position | | *x* | *y* | *z* | | *x* | *y* | *z* |
| 36f | La1 | 0.860467 | 0.021033 | 0.429233 | Ce1 | 0.858200 | 0.016100 | 0.428000 |
| 36f | La2 | 0.663633 | 0.590967 | 0.49606 | Ce2 | 0.664900 | 0.591400 | 0.496100 |
| 18e | La3[a)] | 0.854934 | 0 | ¼ | Ce3 | 0.850700 | 0 | ¼ |
| 36f | C1 | 0.867067 | 0.640533 | 0.293433 | B2 | 0.864300 | 0.637700 | 0.292700 |
| 12c | C2 | 0 | 0 | 0.664733 | B1 | 0 | 0 | 0.665000 |
| 36f | N1 | 0.340833 | 0.458667 | 0.620467 | N2 | 0.336300 | 0.462700 | 0.623700 |
| 36f | N2 | 0.663200 | 0.847800 | 0.413600 | N3 | 0.659000 | 0.850000 | 0.413000 |
| 36f | N3 | 0.295667 | 0.207833 | 0.344433 | N5 | 0.289300 | 0.213700 | 0.345700 |
| 36f | N4 | 0.915067 | 0.897233 | 0.162833 | N4 | 0.908300 | 0.893700 | 0.166700 |
| 6b | N5 | 0 | 0 | 0 | N1 | 0 | 0 | 0 |

[a)] The occupation factor of La3 is refined at 0.853(3).

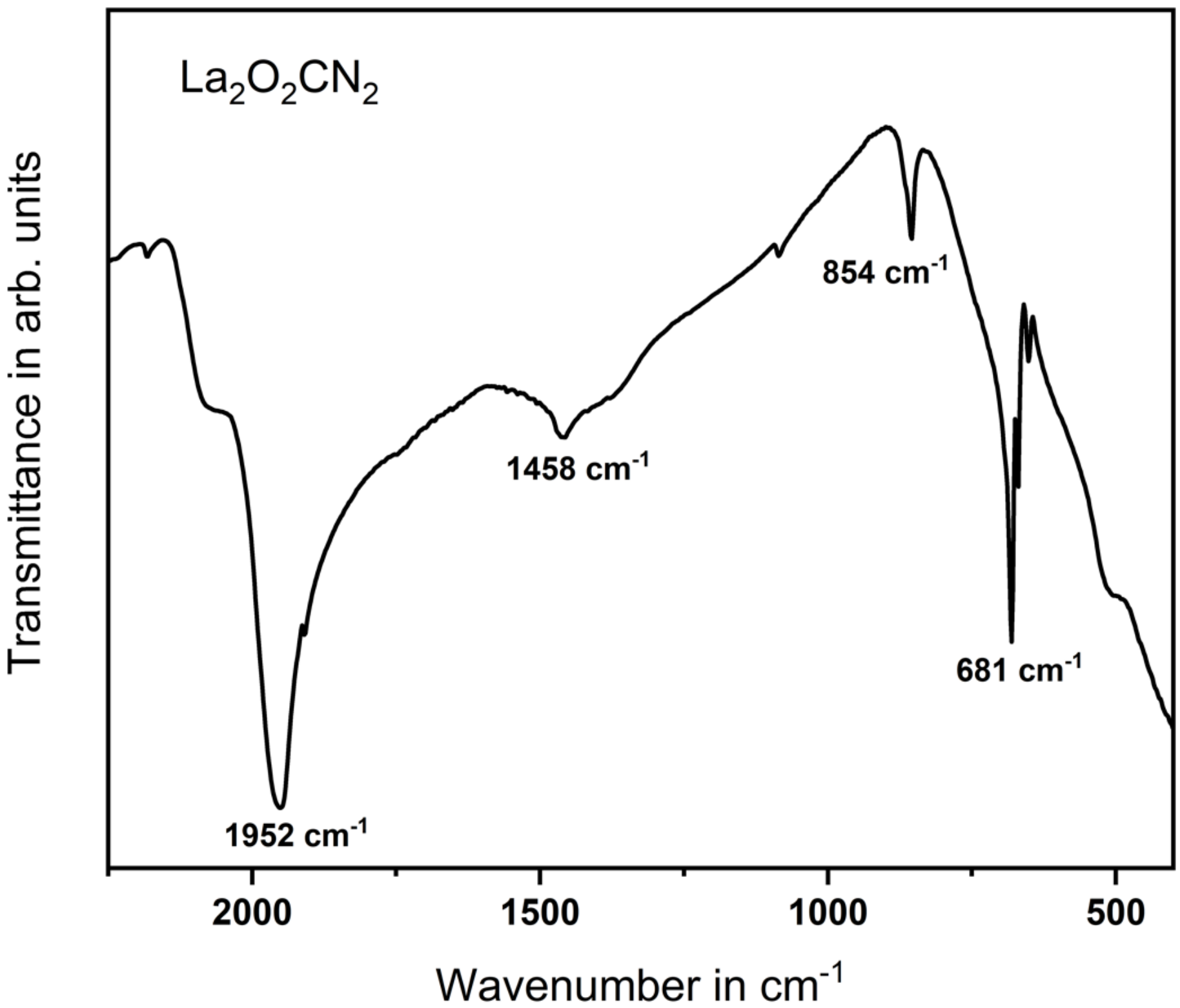

**Figure S1.** IR modes recorded for $La_2O_2CN_2$.

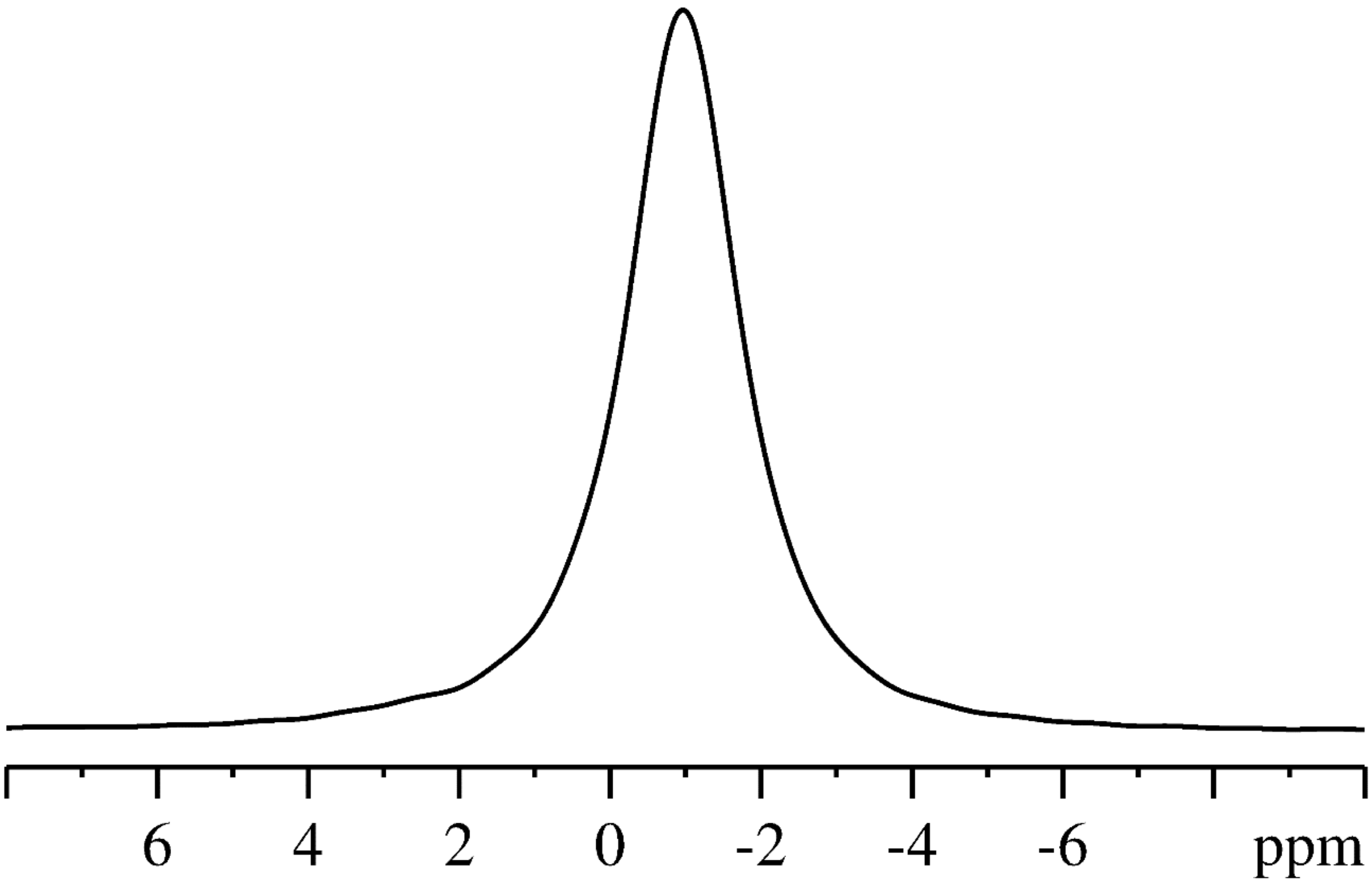

**Figure S2.** $^{7}Li$ MAS NMR of the reaction product powder.

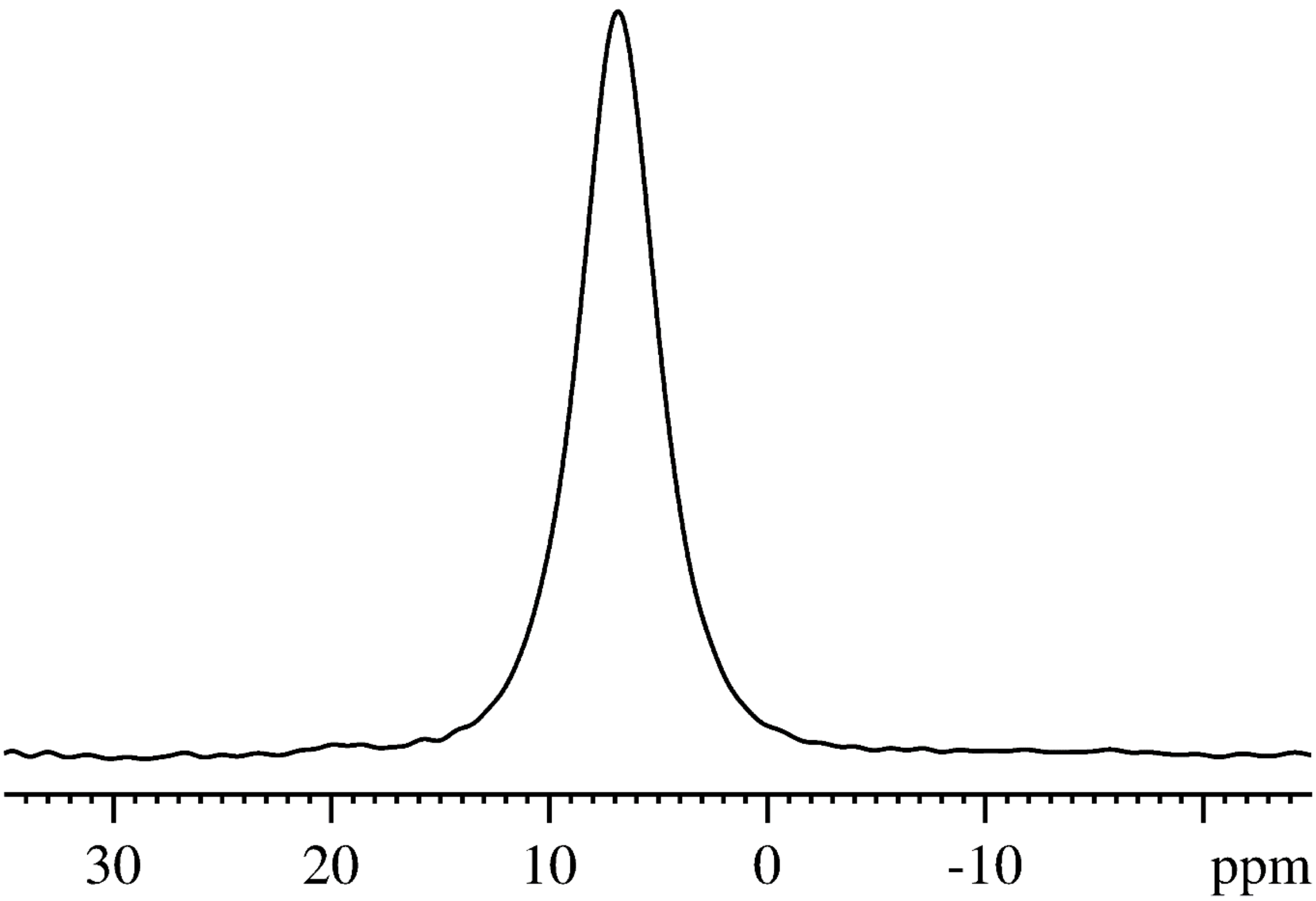

**Figure S3.** $^{23}Na$ MAS NMR of the reaction product powder.

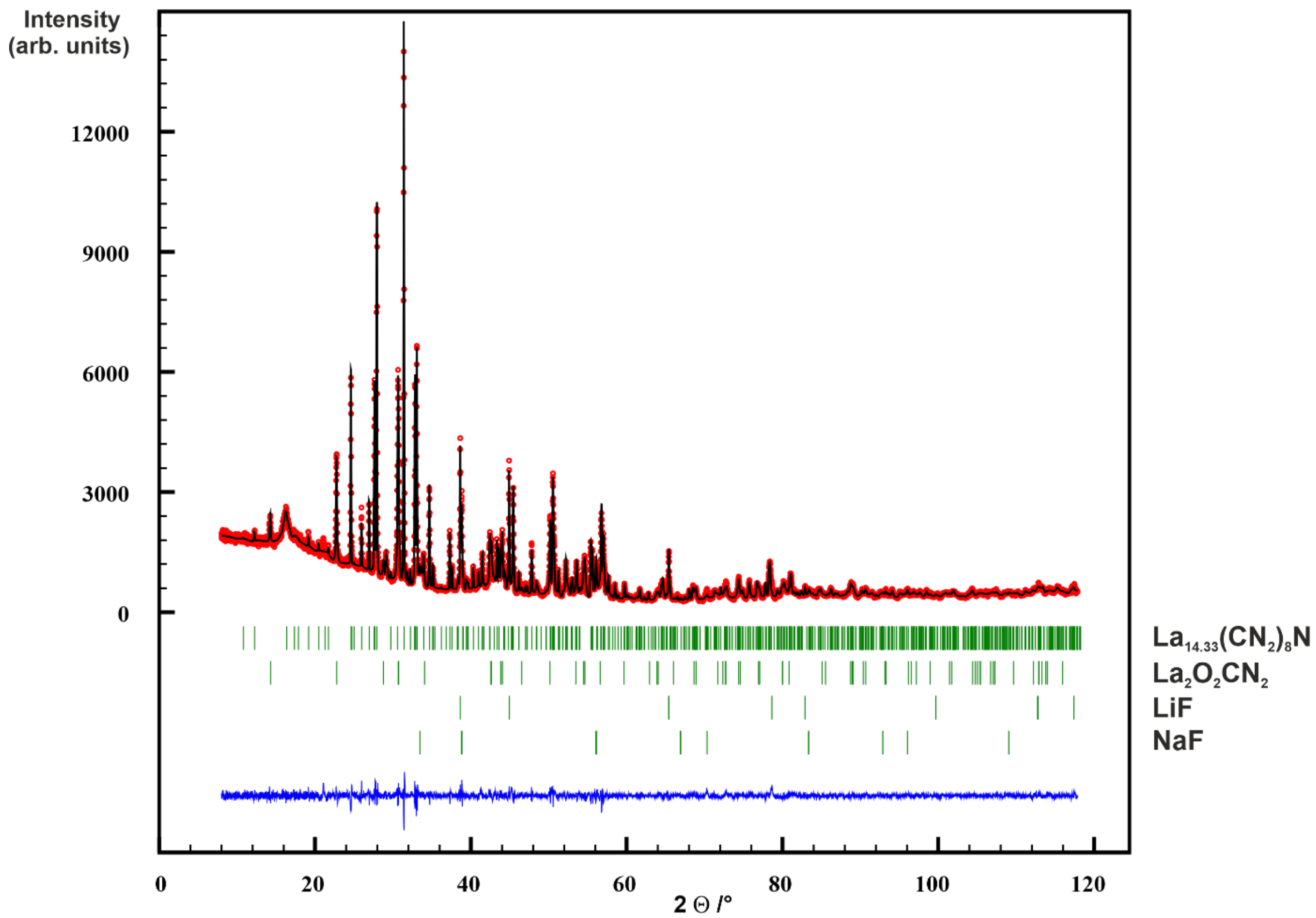

**Figure S4.** Full Rietveld refinement at 298 K of the reaction product powder PXRD. The experimental (red circles) and calculated (black line) intensities, Bragg positions (green lines), and the difference curve (blue lines) are shown.